\documentclass[epj,nopacs]{svjour}
\usepackage{cite}
\usepackage{epsfig}
\usepackage{amssymb}
\begin{document}
\hugehead
\title{Exclusive Leptoproduction of $\rho^0$ Mesons from Hydrogen at 
Intermediate Virtual Photon Energies}

\author{ 
The HERMES Collaboration \\
\ \\ 
A.~Airapetian$^{30}$,
N.~Akopov$^{30}$,
I.~Akushevich$^{7}$,
M.~Amarian$^{23,25,30}$,
J.~Arrington$^{2}$,
E.C.~Aschenauer$^{7,13,23}$,
H.~Avakian$^{11,a}$,
R.~Avakian$^{30}$,
A.~Avetissian$^{30}$,
E.~Avetissian$^{30}$
% ...
} 

\institute{ 
$^1$Department of Physics, University of Alberta, Edmonton, Alberta T6G 2J1, 
Canada$^c$ \\
$^2$Physics Division, Argonne National Laboratory, Argonne, Illinois 
60439-4843, USA$^d$ \\
$^3$Istituto Nazionale di Fisica Nucleare, Sezione di Bari, 70124 Bari, Italy\\
$^4$W.K. Kellogg Radiation Laboratory, California Institute of Technology, 
Pasadena, California 91125, USA$^e$ \\
$^5$Nuclear Physics Laboratory, University of Colorado, Boulder, Colorado 
80309-0446, USA$^f$ \\
$^6$DESY, Deutsches Elektronen Synchrotron, 22603 Hamburg, Germany\\
$^7$DESY Zeuthen, 15738 Zeuthen, Germany\\
$^8$Joint Institute for Nuclear Research, 141980 Dubna, Russia\\
$^9$Physikalisches Institut, Universit\"at Erlangen-N\"urnberg, 91058 Erlangen, 
Germany$^{g,h}$ \\
$^{10}$Istituto Nazionale di Fisica Nucleare, Sezione di Ferrara and Dipartimento di Fisica, Universit\`a di Ferrara, 44100 Ferrara, Italy\\
$^{11}$Istituto Nazionale di Fisica Nucleare, Laboratori Nazionali di Frascati, 00044 Frascati, Italy\\
$^{12}$Fakult\"at f\"ur Physik, Universit\"at Freiburg, 79104 Freiburg, 
Germany$^g$ \\
$^{13}$Department of Subatomic and Radiation Physics, University of Gent, 
9000 Gent, Belgium$^i$ \\
$^{14}$Max-Planck-Institut f\"ur Kernphysik, 69029 Heidelberg, Germany\\
$^{15}$Department of Physics, University of Illinois, Urbana, Illinois 61801, 
USA$^j$ \\
$^{16}$Physics Department, University of Liverpool, Liverpool L69 7ZE, 
United Kingdom$^k$ \\
$^{17}$Department of Physics, University of Wisconsin-Madison, Madison, 
Wisconsin 53706, USA$^l$ \\
$^{18}$Laboratory for Nuclear Science, Massachusetts Institute of Technology, 
Cambridge, Massachusetts 02139, USA$^m$ \\
$^{19}$Randall Laboratory of Physics, University of Michigan, Ann Arbor, 
Michigan 48109-1120, USA$^n$ \\
$^{20}$Lebedev Physical Institute, 117924 Moscow, Russia\\
$^{21}$Sektion Physik, Universit\"at M\"unchen, 85748 Garching, Germany$^g$ \\
$^{22}$Department of Physics, New Mexico State University, Las Cruces, 
New Mexico 88003, USA$^o$ \\
$^{23}$Nationaal Instituut voor Kernfysica en Hoge-Energiefysica (NIKHEF), 
1009 DB Amsterdam, The Netherlands$^p$ \\
$^{24}$Petersburg Nuclear Physics Institute, St. Petersburg, Gatchina, 188350 Russia\\
$^{25}$Istituto Nazionale di Fisica Nucleare, Sezione Sanit\`a and Physics Laboratory, Istituto Superiore di Sanit\`a, 00161 Roma, Italy\\
$^{26}$Department of Physics, Simon Fraser University, Burnaby, 
British Columbia V5A 1S6, Canada$^c$ \\
$^{27}$TRIUMF, Vancouver, British Columbia V6T 2A3, Canada$^c$ \\
$^{28}$Department of Physics, Tokyo Institute of Technology, Tokyo 152, 
Japan$^q$ \\
$^{29}$Department of Physics and Astronomy, Vrije Universiteit, 
1081 HV Amsterdam, The Netherlands$^p$ \\
$^{30}$Yerevan Physics Institute, 375036, Yerevan, Armenia\\
\footnoterule
$ ^a$ supported by INTAS contract No. 93-1827 \\
$ ^b$ partially supported by the Thomas Jefferson National
Accelerator Facility, under DOE contract DE-AC05-84ER40150. \\
$ ^c$ supported by the Natural Sciences and Engineering Research 
Council of Canada (NSERC) \\
$ ^d$ supported by the US Department of Energy, Nuclear Physics Div.,
grant No. W-31-109-ENG-38 \\
$ ^e$ supported by the US National Science Foundation, grant No. PHY-9420470 \\
$ ^f$ supported by the US Department of Energy, Nuclear Physics Div.,
grant No. DE-FG03-95ER40913 \\
$ ^g$ supported by the Deutsche Bundesministerium f\"ur Bildung, 
Wissenschaft, Forschung und Technologie \\
$ ^h$ supported by the Deutsche Forschungsgemeinschaft \\
$ ^i$ supported by the FWO-Flanders, Belgium \\
$ ^j$ supported by the US National Science Foundation, grant No. PHY-9420787 \\
$ ^k$ supported by the U.K. Particle Physics and Astronomy Research Council \\
$ ^l$ supported by the US Department of Energy,  Nuclear Physics Div.,
grant No. DE-FG02-88ER40438, and the US National Science Foundation,
grant No. PHY-9722556 \\
$ ^m$ supported by the US Department of Energy, Nuclear Physics Div. \\
% grant No. ????? \\
$ ^n$ supported by the US National Science Foundation, 
grant No. PHY-9724838 \\
$ ^o$ supported by the US Department of Energy, Nuclear Physics Div.,
grant No. DE-FG03-94ER40847 \\
$ ^p$ supported by the Dutch Foundation for Fundamenteel Onderzoek 
der Materie (FOM) \\
$ ^q$ supported by Monbusho, JSPS and Toray Science Foundation of Japan 
} 

\date{Received: April 25, 2000; Revised: July 3, 2000}
\titlerunning{Diffractive Electroproduction of $\rho^0$ Mesons}
\authorrunning{The HERMES Collaboration}

\abstract{
Measurements of the cross section for exclusive virtual-photoproduction
of $\rho^0$ mesons 
from hydrogen are reported. The data were collected by the HERMES experiment
using 27.5~GeV positrons incident on a hydrogen
gas target in the HERA storage ring. The invariant mass 
$W$ of the photon-nucleon system
ranges from 4.0 to 6.0 GeV, while the negative squared four-momentum $Q^2$
of the virtual photon varies from 0.7 to 5.0 GeV$^2$. The 
present data together with most of the previous data in the
intermediate $W$-domain
are well described by a model that infers the $W$-dependence of the 
cross section from the dependence on the Bjorken scaling variable $x$
of the unpolarized structure function for deep-inelastic scattering.
In addition, a model calculation 
based on Off-Forward Parton Distributions gives a fairly good account of the 
longitudinal component of the $\rho^0$ production cross section for 
$Q^2$ $>$ 2 GeV$^2$.
}

\maketitle

%%%%%%%%%%%%%%%%%%%%%%%%%%%%%%%%%%%%%%%%%%%%%%%%%%%%%%%%%%%%%%%%%%%%%%%%%%%%%
\section{Introduction}\label{sec:intro}

This paper presents cross section measurements for
exclusive diffractive production of $\rho^0$(770) vector mesons in
positron scattering on a $^1$H target.
The production of vector mesons by real or virtual photons
is of considerable interest, as the corresponding
cross section is closely related to other observables
in lepton scattering. For example,
with help of the recently introduced Off-Forward Parton
Distributions (OFPDs), one can relate elastic nucleon form 
factors, deep-inelastic scattering (DIS) structure functions, 
virtual Compton scattering cross sections, and vector meson 
production cross sections~\cite{Rady,Ji,Collins,FS96,MvdH,MvdH2,Mank1,Mank2}. 
The OFPDs
represent a generalization of the parton distributions measured,
for instance, 
in inclusive DIS experiments. 

In Refs.~\cite{MvdH,MvdH2} the longitudinal part of the $\rho^0$
virtual-photoproduction cross section is calculated in the
OFPD framework. At large values of the
photon-nucleon invariant mass ($W >$ 10 GeV) the calculated
cross section is dominated by a two-gluon 
exchange mechanism, which has been treated using
the perturbative approach of Ref.~\cite{FS96}. These
calculations reproduce existing data. 
However, uncertainties arise due to
the size of higher-order and higher-twist
contributions~\cite{FS96,Mank1,Mank2,Beli99}, which
are larger at small values of $Q^2$
(the negative square of the four-momentum of
the virtual photon).
Below 10 GeV the calculated cross section is dominated
(in leading twist) by a handbag diagram, in which the virtual photon
is absorbed by a valence quark in the nucleon~\cite{MvdH,MvdH2}.
Following Ref.~\cite{MvdH} this mechanism is called
quark exchange. 
However, few data are available between 4
and 10 GeV.
Several data sets exist for $W <$ 4 GeV, but in this 
domain the OFPD calculations do not apply, as the reaction 
receives contributions from other reaction channels
in this domain.

The relation between the vector meson leptoproduction
cross section and the structure function $F_2^{\mathrm p}(x)$ of the proton
(with $x$ the Bjorken scaling variable) is apparent 
in a model calculation by Haakman et al.~\cite{ref4} 
based on Reggeon field theory.
In this model the $x$-dependence of $F_2^{\mathrm p}(x)$ is related to the
$W$-dependence of the cross section for vector meson leptoproduction.
The model gives a good description of
$\gamma^* p \rightarrow \rho^0 p$ cross sections
for $W$ ranging from 10 to 180 GeV.
Between 4 and 10 GeV the calculated shape of the
$W$-dependence changes significantly with $Q^2$.
The model has not yet been properly assessed in
this domain due to lack of data.

The two examples above illustrate the need for additional
exclusive $\rho^0$ leptoproduction data in the $W$-range
between 4 and 10 GeV. Such data are reported in the present 
paper. Exclusive cross sections have been measured for $W$ 
values between 4.0 and 6.0 GeV,
and $Q^2$ values between 0.7 and 5.0 GeV$^2$.
The kinematical variables used to describe $\rho^0$ leptoproduction
are introduced in the next section. Experimental details are
provided in section 3, while the data analysis is discussed
in section 4. The experimental results are compared to
various model calculations in section 5, and the paper is
summarized in the last section.

%%%%%%%%%%%%%%%%%%%%%%%%%%%%%%%%%%%%%%%%%%%%%%%%%%%%%%%%%%%%%%%%%%%%%%%%%%%

\section{Kinematics}\label{sec:kinem}

In the present leptoproduction measurement
both the scattered lepton and two oppositely charged hadrons
are observed. 
The $\rho^0$ leptoproduction events are identified by
requiring that the reconstructed invariant mass
of the two hadrons is close to 
the mass of the $\rho^0$ meson, i.e. 0.77 GeV.
The kinematic variables that characterize such
measurements are defined here in the laboratory frame.
The variables describing the kinematics of the
virtual photon include its energy $\nu$,
its fractional energy $y = \nu / E_0$ (with $E_0$ 
the incident lepton energy), and its four-momentum $q$.
The latter quantity is related to the previously introduced 
variable $Q^2$ through $Q^2 = -q^2 > 0$. The Bjorken scaling
variable $x$ is related to $\nu$ and $Q^2$ by $x = Q^2/(2 M \nu)$,
and the photon-nucleon invariant mass $W$ is given 
by $W^2 = M^2 + 2 M \nu - Q^2$.
The virtual-photon polarization parameter
is represented by $\epsilon$.
 
The $\rho^0$ meson is characterized by 
its four-momentum $v$ and $\theta_{\pi\pi}$,
which represents the angle between the two pions
into which the $\rho^0$ meson decays. 

Combining some of the lepton and meson
variables, one can introduce $t = (q - v)^2 < 0$,
the square of the four-momentum exchange between the virtual 
photon and the target,
and $\Delta E = (P_{\mathrm Y}^2 - M^2)/2M$, which is a measure of the
missing energy (where $P_{\mathrm Y} = P + q - v$ represents the 4-momentum
of the unobserved final state $\mathrm Y$ with $P = (M,0)$ that of the
target nucleon). 

Instead of $-t$, the above-threshold momentum
transfer $-t' = -t + t_0$ is often used, which is approximately
equal to $p_t^2$ -- the square of the transverse momentum of 
the $\rho^0$ meson
with respect to the direction $\vec{q}$. In this expression 
$-t_0$ represents the minimum value of $-t$ for fixed values of
$\nu$, $Q^2$ and $P_{\mathrm Y}^2$.

%%%%%%%%%%%%%%%%%%%%%%%%%%%%%%%%%%%%%%%%%%%%%%%%%%%%%%%%%%%%%%%%%%%%%%%%%

\section{Experiment}\label{sec:data}

The data were collected during the 1996 and 1997 running periods
of the HERMES experiment~\cite{specpaper} at DESY using a 27.5 GeV
longitudinally polarized positron beam with a $^1$H gas target
in the HERA storage ring.
Part of the data set was collected with
longitudinally polarized targets. Since the polarization degrees
of freedom were not exploited in the present analysis,
the average over both target polarization states is taken.

The HERMES polarized proton target~\cite{ABStarget} is formed
by injecting a nuclear-polarized beam of atomic hydrogen
from an atomic beam source into a tubular open-ended
storage cell inside the positron ring.
The cell provides a 40 cm long target of pure
atomic species with an areal density
of approximately 7 $\times$ 10$^{13}$ atoms/cm$^2$. 
With unpolarized targets of molecular hydrogen, 
areal densities of about 10$^{15}$ atoms/cm$^2$ were obtained.
The storage cell was shielded from synchrotron radiation by two
sets of collimators, one of which is 
moveable. No particles were observed to originate
from scattering events in the walls of the storage cell. 

During the course of one fill (typically 8 hours long), the positron current
in the ring decreased from typically 30 -- 40 mA at injection
to $\sim 10$ mA, at which point the ring was emptied. 
The data presented in this paper
correspond to an integrated luminosity
of 108 pb$^{-1}$. 

The HERMES spectrometer is described in detail elsewhere~\cite{specpaper}.
It is a forward spectrometer in which both the scattered positron
and produced hadrons are detected within an
angular acceptance $\pm$ 170 mrad horizontally, and $\pm$ (40 -- 140) mrad
vertically.
The scattered-positron trigger was formed from a coincidence between 
a pair of scintillator
hodoscope planes and a lead-glass calorimeter.
The trigger required an energy of more than 3.5 GeV deposited in the
calorimeter. (For part of the running the trigger threshold
was reduced to 1.5 GeV, for which a correction was applied
in the data analysis.)
Positron identification was accomplished using the calorimeter, the
preshower counter consisting of the second hodoscope preceded by
a lead sheet, a transition-radiation detector,
and a threshold gas $\breve {\mathrm C}$erenkov counter. This system provided
positron identification with an average efficiency of 99\% and a
hadron contamination of less than 1\%.

%%%%%%%%%%%%%%%%%%%%%%%%%%%%%%%%%%%%%%%%%%%%%%%%%%%%%%%%%%%%%%%%%%%%%%%%%

\section{Data analysis}

Only those events were selected
that contained a scattered positron and exactly
two hadrons with opposite charge. (A more detailed description
of the analysis is given in Refs.~\cite{Machiel,belz}.)
A number of geometric requirements were
imposed on the particle tracks to ensure that they were well contained
within the acceptance of the spectrometer.
It was also required that the tracks
originated from along the beam line within $\pm$ 18 cm
of the centre of the target. In addition, several constraints
were imposed on the kinematic variables.
The size of the radiative corrections was limited by
requiring $y \le$ 0.85. Because the $W$-acceptance
of the HERMES spectrometer for $\rho^0$ production is 
sharply reduced both below 4 GeV and above 6 GeV, cross
sections for only two $W$-bins (4--5 and 5--6 GeV) have been
extracted from the data. 

The $\rho^0$ vector mesons were identified by
requiring 0.6 GeV $< M_{\pi \pi} <$ 1 GeV, with $M_{\pi \pi}$ the
invariant mass of the pair of detected hadrons, assuming that they are pions.
It has been verified that this requirement also removes the 
$\phi \rightarrow \mathrm{K^{+} K^{-}}$ background,
by confirming that the $\phi$ events, which occur at $M_{\mathrm K \mathrm K}$
$\approx$ $M_{\phi}$, appear in the $M_{\pi \pi}$ spectrum at
$M_{\pi \pi} <$ 0.6 GeV. Here 
$M_{\mathrm K \mathrm K}$ is the invariant mass calculated 
assuming that the two hadrons are kaons, and $M_{\phi}$ = 1.019 GeV is
the mass of the $\phi$ meson.

\begin{figure}[t]
\begin{center}
\includegraphics[width=0.47\textwidth]{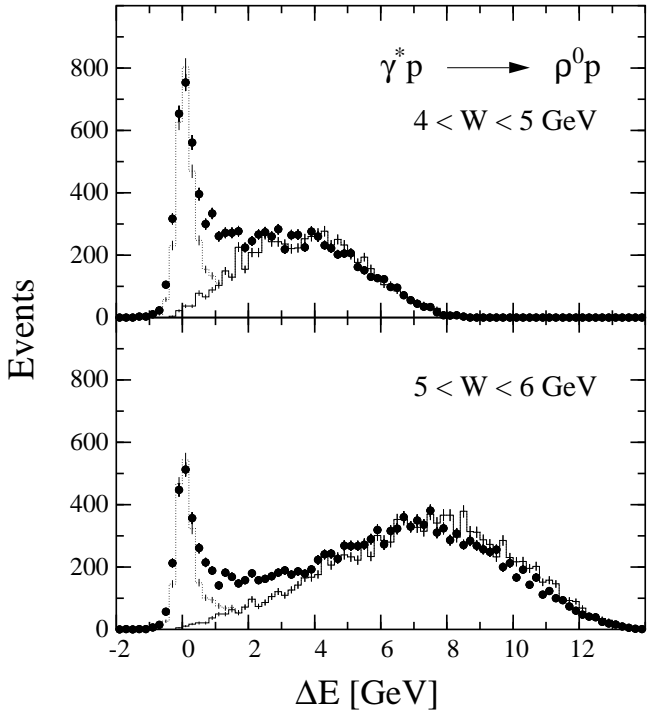}
\end{center}
\caption[dummy]{
Missing energy spectra for leptoproduction of
$\rho^0$ mesons off $^1$H at an incident energy of 27.5 GeV. The
upper (lower) panel corresponds to the $W$-range between 4 and
5 (5 and 6) GeV. The data have been selected requiring
0.6 GeV $< M_{\pi\pi} <$ 1 GeV and $-t' <$ 0.4 GeV$^2$.
The solid histograms represent the results of a Monte Carlo 
simulation~\cite{LEPTO} of the background, which has been scaled to 
the data for $\Delta E >$ 3.0 GeV. The dotted histogram includes
a simulation~\cite{belz} of exclusive $\rho^0$-production as well.
As the Monte Carlo simulations do not include inelastic nucleon
excitations or internal radiative effects for the exclusive
channel, the $\Delta E$-region between
0.5 and 2.5 GeV cannot be properly described.
}
\label{fig:deltaE}
\end{figure}

In exclusive $\rho^0$ electroproduction, 
i.e. $e N \rightarrow e' \rho^0 N$,
the part of the final state that is unobserved
at HERMES consists of a 
nucleon recoiling without excitation. Such events 
are selected by requiring that
the missing energy $\Delta E$ is approximately 
zero. In this domain the $\Delta E$ spectrum,
as displayed in Fig.~\ref{fig:deltaE} for the two $W$-bins,  
shows a clear peak near $\Delta E \approx$ 0.
The exclusive events were selected
by requiring $\Delta E$ $<$ 0.4 GeV. 

Using the Monte Carlo simulation for exclusive $\rho^0$
production described in section 4.2, it has been
evaluated that on average 18\% of the exclusive
events fall outside the imposed $\Delta E$ window.
This effect has been accounted for (bin by bin)
in the acceptance correction.

The requirement on $\Delta E$ is relatively tight
in order to suppress inelastic contributions
involving nucleon excitations. As no Monte Carlo
simulation is available for such double diffractive
processes, their remaining contribution is
estimated from other data in section 4.3.

%%%%%%%%%%%%%%%%%%%%%%%%%%%%%%%%%%%%%%%%%%%%%%%%%%%%%%%%%%%%%%%%%%%%%%%%

\subsection{Background subtraction}

There is a background contribution under the exclusive
$\rho^0$ peak that
is caused by hadrons from DIS fragmentation processes. Part
of this background is removed by excluding the region $-t'$ $>$ 0.4 GeV$^2$,
where the background dominates the $\rho^0$ yield. The
remaining background is estimated and subtracted using the
methods described below.

The background estimate is based on the LEPTO Monte Carlo
program~\cite{LEPTO}. This simulation includes hadrons resulting 
from fragmentation processes in deep-inelastic scattering (using the
Lund fragmentation code),
but not from diffractive $\rho^0$ production.
The $\Delta E$ spectrum was generated independently for each
($Q^2$,$W$)-bin, and normalized to the experimental data in the
$\Delta E$ region above 3.0 GeV (see Fig.~\ref{fig:deltaE}).
Thus normalized, the background was
subtracted from the data in the $\Delta E$ domain that was
used for the determination of the cross section. Depending on
kinematics the background contribution ranges from 
(3 $\pm$ 1)\% to (9 $\pm$ 4)\%, where the uncertainties
include the effect of varying the $\Delta E$-ranges used
in evaluating the background contribution. The lower limit of
the $\Delta E$-range used to normalize the background was varied
from 3 to 5 GeV, and  the upper limit used in selecting
the exclusive events was varied from 0.3 to 0.6 GeV.

The background subtraction procedure was verified by also using
an alternative method. In this case a $\Delta E$ spectrum was 
extracted from the data by requiring $-t'$ values 
between 0.7 and 5 GeV$^2$, i.e. well beyond the region 
used to determine the diffractive cross section.
A similar normalization procedure is used, i.e. the
background spectrum is normalized such as to reproduce
the same number of events for $\Delta E >$ 3 GeV as in the spectrum
obtained with the standard $-t'$ requirement. The tail of the
normalized background spectrum for $\Delta E <$ 0.4 GeV
was taken to represent the actual background contribution
to the $\rho^0$ mass peak. The cross sections obtained after this
background subtraction method were found to be consistent with
those obtained using the Monte-Carlo based method within 
the systematic errors listed below in Table~\ref{tab:cross}.

%%%%%%%%%%%%%%%%%%%%%%%%%%%%%%%%%%%%%%%%%%%%%%%%%%%%%%%%%%%%%%%%%%%%%%%%%

\subsection{Acceptance correction}

The data were corrected for the finite acceptance and
inefficiencies of the
HERMES spectrometer for $\rho^0$ production. The correction was
evaluated using a Monte Carlo simulation based on the
Vector Meson Dominance (VMD) model. Details are given 
in Ref.~\cite{belz}.
For each kinematic variable ($\nu$, $Q^2$, $W$, $M_{\pi\pi}$,
$\cos{\theta_{\pi\pi}}$ and $-t'$) good agreement between the measured
distribution and the simulation was obtained.

The acceptance correction is large because the HERMES spectrometer
has a relatively small acceptance for $\gamma^* p \rightarrow
\rho^0 p$ events. The acceptance for $\rho^0$ mesons (given an
observed scattered positron) ranges from only
5.9\% at low $Q^2$ to 18\% at high $Q^2$. As the corresponding
correction factors are large, it is mandatory to study the 
dependence of the acceptance correction
on the details of the Monte Carlo calculation. For that purpose,
the entire analysis was redone varying the assumptions for the 
angular distributions of the production and decay of the $\rho^0$-meson.
The original Monte Carlo simulation, which assumed
s-Channel Helicity Conservation (SCHC), was replaced 
by one that produces uniform angular distributions for all relevant
angles.
In addition, the $Q^2$-dependence of
the VMD propagator (as given by Eq.~\ref{eq:res-fittingf} in
section 5.1) was varied by changing the exponent from 2 to 2.5,
and an alternative event generator DIPSI~\cite{dipsi} was used.
The sensitivity of the Monte Carlo yield to variations of the angular
distributions and changes of the assumed $Q^2$-dependence leads to
typical systematic uncertainties of 10\% and 6\%, respectively.
The effect of the event generator itself amounts to typically 8\%.
Including all variations, the uncertainty of the cross sections
due to the acceptance correction ranges from 12\% to 17\%,
depending on the kinematics. 

%%%%%%%%%%%%%%%%%%%%%%%%%%%%%%%%%%%%%%%%%%%%%%%%%%%%%%%%%%%%%%%%%%%%%%

\subsection{Cross section determination}

\begin{figure}[t]
\begin{center}
\includegraphics[width=0.47\textwidth]{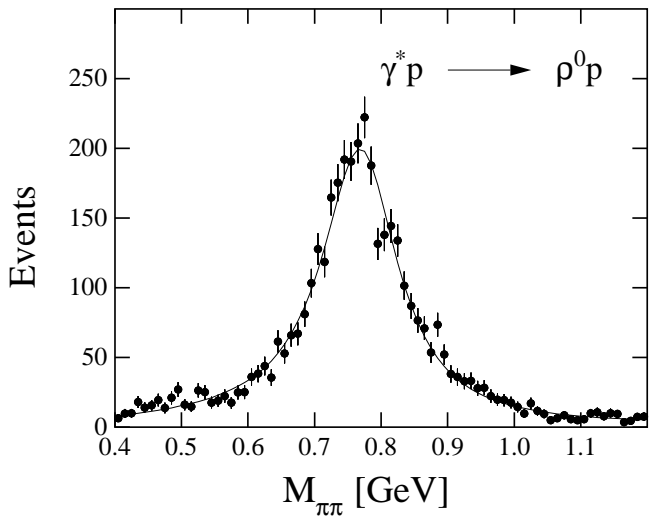}
\end{center}
\caption[dummy]{
Invariant mass spectrum of the two oppositely charged
hadrons assuming that they are pions. Backgrounds have
been subtracted, the contribution from $\phi$ decays has been removed
by requiring that $M_{\mathrm K \mathrm K}$ $>$ 1.04 GeV,
and the acceptance correction has been
applied. The $\rho^0$ mass peak at 770 MeV is
described by a Breit-Wigner function (solid line), of 
which the parameters were fitted to the data.
}
\label{fig:Mpipi}
\end{figure}

The $M_{\pi\pi}$ mass spectrum is shown in Fig.~\ref{fig:Mpipi}. In this
spectrum backgrounds are subtracted, the acceptance correction
is applied and all kinematic requirements mentioned above are
imposed. The $\rho^0$ mass peak was fitted using a Breit-Wigner (p-wave) 
function and a constant background term representing the non-resonant 
physics background. The Breit-Wigner function was multiplied
by the factor $(M_{\rho}/M_{\pi\pi})^{n_s}$ in which $n_s$ represents
the skewing parameter. This procedure is
identical to the one described in Ref.~\cite{e665}, and
was first proposed by Ross and Stodolsky~\cite{RS}.
The resulting fit parameters
($M_{\rho^0}$ = 0.776 $\pm$ 0.003 GeV,
$\Gamma_{\rho^0}$ = 0.147 $\pm$ 0.009 GeV, and 
$n_s$ = 2.2 $\pm$ 0.4) are in good agreement with the values
published by E665~\cite{e665}: $M_{\rho^0}$ = 0.777 $\pm$ 0.002 GeV,
$\Gamma_{\rho^0}$ = 0.146 $\pm$ 0.003 GeV, 
and $n_s$ = 2.7 $\pm$ 0.6 (for the same $Q^2$-region). Both
experiments find a background term that is consistent with
zero, and are in fair agreement with the PDG-values~\cite{booklet}:
$M_{\rho^0}$ = 0.7700 $\pm$ 0.0008 GeV, 
$\Gamma_{\rho^0}$ = 0.1507 $\pm$ 0.0011 GeV.
If the effect of the $\omega$ decay is also considered
in the fit, $\omega$ mass and width parameters are found that
are also consistent with Ref.~\cite{booklet}. The
contribution of this interference averages out to zero in the
considered mass interval. The absolute contribution of
$\omega$-production to the apparent $\rho^0$ yield is estimated
to be less than 1\%, and has therefore been neglected.
It should be noted that the $\rho^0$ production cross sections
presented in this paper correspond to the data in the mass region
0.6 GeV $< M_{\pi\pi} <$ 1.0 GeV with a correction (of 14 $\pm$ 2\%)
for the tails of the skewed Breit-Wigner distribution~\cite{RS}
outside this mass window. This correction factor has been
evaluated relative to the mass range between $2 M_{\pi}$ and
$(M_{\rho^0} + 5 \Gamma_{\rho^0})$.

Examples of the $-t'$ distribution
of the data can be found in Refs.~\cite{belz,ref7}. 
The $-t'$ distribution is fit to a falling
exponential, and the fit is used to correct the measured
cross section for the excluded region $-t' >$ 0.4 GeV$^2$.
The slope parameter $b$ from the fit is (6.82 $\pm$ 0.15) GeV$^{-2}$,
which is consistent with previously published values.

The $\rho^0$ production data were normalized using the inclusive
deep-inelastic scattering (DIS) yield derived from the same data
sample. By comparing the DIS yield
to the LEPTO Monte-Carlo yield based on the
world data~\cite{ref6} evaluated at the
appropriate $Q^2$ and $x$ values,
the required normalization factors were determined.
Virtual photoproduction cross sections, $\sigma_{\gamma^* p \rightarrow
\rho^0 p}$, were obtained from the leptoproduction
cross section after applying the photon flux factor $\Gamma_T$
(using the Hand convention as outlined in Ref.~\cite{e665}, for instance):
\begin{equation}
\sigma_{\gamma^* p \rightarrow \rho^0 p} = { \frac{ 1 } {\Gamma_T}
  \frac{d \sigma_{e p \rightarrow e' \rho^0 p}} {d \nu dQ^2} }
\label{eq:Xsect}
\end{equation}

In double-diffractive (DD) processes the target nucleon is broken up.
In a recent study of target dissociation in $\rho^0$ production
at center-of-mass energies between 60 and 180 GeV, the cross
section ratio
of double-diffractive to single-diffractive $\rho^0$ production
was measured to be 0.65 $\pm$ 0.17~\cite{H1}. Accounting for
the shape of the baryonic spectrum (taken from $\mathrm p \mathrm p$
diffractive scattering~\cite{Akimov}) and the resolution of the present
experiment, the DD contribution to the
diffractive $\rho^0$ production cross section at HERMES
is found to be (4 $\pm$ 2)\%, for which the data were corrected.
The uncertainty in the DD contribution originates from a
rough estimate of the acceptance for double diffractive events
relative to that for single diffractive $\rho^0$ events, and the
uncertainty of the data of Ref.~\cite{H1}. 
Taking into account other
estimates of the ratio of double diffractive to single 
diffractive $\rho^0$ production also at lower $W$-values~\cite{Holt},
which are smaller in general, a total systematic uncertainty of about
3\% (relative to the measured cross section) has been assigned 
to the DD contribution.
The small size of the DD contribution is related to the
good energy resolution of the HERMES experiment ($\sigma_{\Delta E}
\approx$ 0.25 GeV -- see Fig.~\ref{fig:deltaE}).

The final cross sections are obtained by
applying a radiative correction to the experimentally determined
$\rho^0$ production cross section. The internal radiative
correction has been evaluated separately for each ($Q^2$,$W$) bin,
and typically amounts to 18\%~\cite{radcor}. External radiative
effects (caused by detector materials) are included in the
acceptance Monte Carlo.

%%%%%%%%%%%%%%%%%%%%%%%%%%%%%%%%%%%%%%%%%%%%%%%%%%%%%%%%%%%%%%%%%%%%%%%%

\section{Results}

In Table~\ref{tab:cross} the results for the $\rho^0$
virtual-photoproduction cross section are given for
each $Q^2$ and $W$ bin.
The systematic uncertainties are dominated by those from
the acceptance correction factors, which amount to
17\%, 11\%, 14\% and 17\% in the bins centered at $Q^2$ values
of 0.83, 1.3, 2.3 and 4.0 GeV$^2$, respectively. Another important 
contribution to the systematic uncertainty stems from from the
uncertainty in the reconstruction and data selection efficiency,
which totals about 9\%. This contribution has
been estimated by varying the large number
of requirements used to select data and
reconstruct valid events. The quoted 9\% also
includes uncertainties in tracking effeciency,
kaon contamination, radiative corrections and
DD-contribution. Remaining contributions to the total
systematic uncertainty are the absolute 
normalization (6\%) and the background
subtraction (4\%). The combined systematic
uncertainty on the $\rho^0$ cross sections ranges
from 16 to 21\%, depending on the kinematics.

\begin{table} [bt]
\caption{The measured virtual-photoproduction cross sections
$\sigma_{\gamma^* p \rightarrow \rho^0 p}$ for exclusive $\rho^0$
production on $^1$H (in $\mu$b) corrected for radiative effects. Both
the statistical (first) and systematic (second) uncertainties
are listed.
\label{tab:cross}
}
\begin{center}
\begin{tabular}{|c|c|c|}
\hline
$\langle Q^2 \rangle$ [GeV$^2$] &  $\langle W \rangle$ = 4.6 GeV &
$\langle W \rangle$ = 5.4 GeV \\
\hline
 0.83  & 2.46 $\pm$ 0.13 $\pm$ 0.51   &  2.04 $\pm$ 0.10 $\pm$ 0.43 \\
 1.3  & 0.92 $\pm$ 0.03 $\pm$ 0.15   &  1.00 $\pm$ 0.04 $\pm$ 0.16 \\
 2.3  & 0.43 $\pm$ 0.02 $\pm$ 0.08   &  0.41 $\pm$ 0.02 $\pm$ 0.07 \\
 4.0  & 0.16 $\pm$ 0.01 $\pm$ 0.03   &  0.10 $\pm$ 0.01 $\pm$ 0.02 \\
\hline
\end{tabular}
\end{center}
\end{table}

\begin{figure} [t]
\begin{center}
\includegraphics[width=0.47\textwidth]{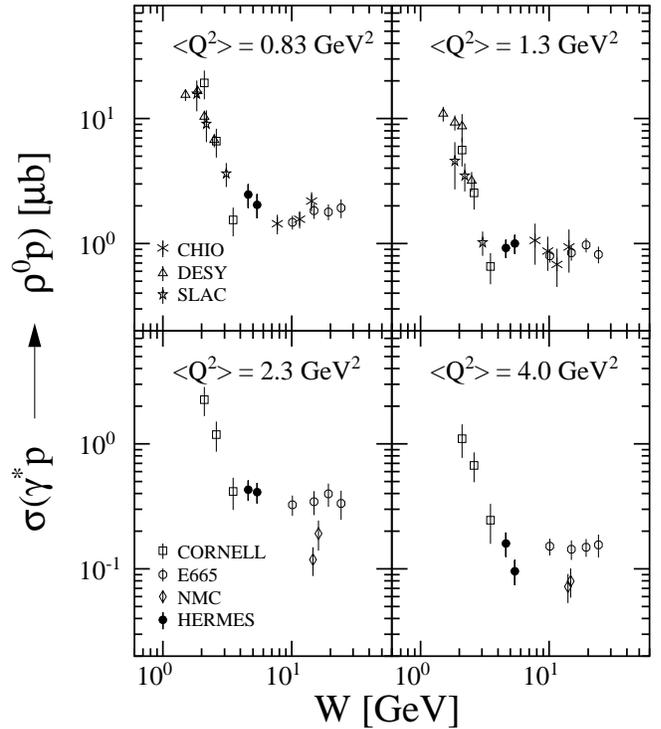}
\caption[]{\label{fig:crossw1}
The virtual-photoproduction cross section for exclusive $\rho^0$
production versus $W$ for the indicated values of
$\langle Q^2 \rangle$.
The data collected in the
present experiment are represented by solid circles.
The open squares are from Ref.~\cite{Cassel}, the open triangles from
Ref.~\cite{Joos}, the open stars from Ref.~\cite{Papa}, the
crosses from Ref.~\cite{Sham}, the open circles
from Ref.~\cite{e665}, and the open diamonds from
Ref.~\cite{nmc}. Previous data have been scaled to the
presently used $Q^2$-bins using Eq. (2). The error bars
include both the statistical
and systematic uncertainties (added in quadrature).
}
\end{center}
\end{figure}

In Fig.~\ref{fig:crossw1} the virtual-photoproduction 
cross section is plotted versus $W$, and compared to existing
measurements \cite{e665,Joos,Papa,Cassel,Sham,nmc} at nearby
values of $W$. As the various data sets have been measured at 
different average $Q^2$ values, most data sets were
rescaled to the average $Q^2$ values of the HERMES data
using the $Q^2$-dependence of the VMD model, which is in 
agreement with the present data as will be shown below 
(see Fig.~\ref{fig:crossqq2} and Eq.~(\ref{eq:res-fittingf})
with $m$ = 2). It is noted that the various data sets have
also been obtained at different average $\epsilon$-values,
for which no correction has been applied.
The HERMES data are seen to fill the gap which
previously existed between the steeply declining data at low $W$, and the
data collected at higher $W$ values, which have a much
flatter $W$-dependence.
With a few exceptions, there is a fair consistency
among the world's data on exclusive $\rho^0$ production.
The first exception concerns the highest $W$ point (at 3.5 GeV)
of the Cornell data~\cite{Cassel} at $Q^2$ = 0.83 GeV$^2$, which
deviates by about 2$\sigma$--3$\sigma$ from
a smooth interpolation of all other data in that $W$-region.
Secondly, there is a difference of about a factor
of two between the NMC~\cite{nmc} and E665 data~\cite{e665}
at $W \approx$ 15 GeV in the two highest $Q^2$-bins. The latter 
discrepancy has been 
reported before~\cite{e665,zeus}. Part of the discrepancy
might be related to a (model-dependent) subtraction of the
non-resonant contribution to the $\rho^0$-peak that was
applied by NMC~\cite{nmc} in contrast to E665~\cite{e665} and
the present experiment.

It should be noted that it is the total cross section for
the process $\rm{ep} \rightarrow \rm{eh^+h^-p}$
with $M_{\pi \pi} \approx M_{\rho^0}$ that is displayed in 
fig.~\ref{fig:crossw1} for all data sets. This total cross
section is believed to receive contributions from exclusive $\rho^0$ production
through Reggeon exchange, as well as from reaction channels such
as those involving nucleon resonances at low $W$, as reported 
in Refs.~\cite{Joos,Papa,Cassel}.
In virtual-photoproduction it is usually assumed that the
latter  contributions 
are negligible beyond $W \approx$ 4 GeV.
The data of Fig.~\ref{fig:crossw1} indicate that these additional processes
have a decreasing contribution to $\rho^0$ production
up to about 4--5 GeV,
where the steep decrease of the cross section changes into an almost
flat $W$-dependence. Henceforth, we restrict the comparison with
existing calculations to data collected at $W >$ 4 GeV.

%%%%%%%%%%%%%%%%%%%%%%%%%%%%%%%%%%%%%%%%%%%%%%%%%%%%%%%%%%%%%%%%%%%%%%%%%%%

\subsection{$Q^2$-dependence}

\begin{figure} [t]
\begin{center}
\includegraphics[width=0.47\textwidth]{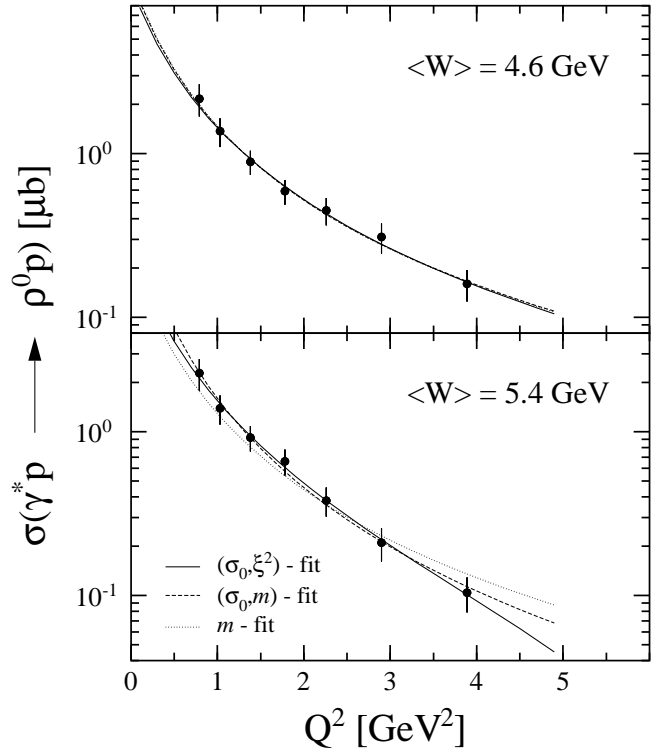}
\caption[]{
The virtual-photoproduction cross section for $\rho^0$
production on $^1$H versus $Q^2$ for \mbox{4 GeV $< W <$ 5 GeV}
(upper panel) and \mbox{5 GeV $< W <$ 6 GeV} (lower panel).
The error bars include both the statistical and systematic uncertainties,
added in quadrature.
The solid, dashed and dotted curves represent fits using the values
of $\sigma_0$ and $\xi^2$, $\sigma_0$ and $m$, or only $m$
as free parameters, respectively.
\label{fig:crossqq2}
}
\end{center}
\end{figure}

The HERMES data are displayed in Fig.~\ref{fig:crossqq2} 
as a function of $Q^2$ for the two $W$ bins.
For the purpose of extrapolating to $Q^2$ = 0, the $Q^2$-dependence 
of the data has been parameterized using the following
functional form, inspired by the VMD model~\cite{e665,Bauer}
\begin{equation}
  \sigma(Q^{2}) = \sigma_0 \cdot
    \left( \frac{ M_{\rho}^2 }{ Q^2 + M_{\rho}^2 } \right)^m \cdot
      ( 1 + \epsilon R(Q^2)).
\label{eq:res-fittingf}
\end{equation}
The VMD-model predicts that $m$ = 2 and that the ratio of
longitudinal to transverse $\rho^0$ photoproduction cross sections
is given by $R = { \xi^2 Q^2 } / { M_{\rho}^2 }$.
A good description of the data is obtained by fixing $m$
to its VMD value, and
treating $\sigma_0$ and $\xi^2$ as free parameters. Here $\sigma_0$
represents the $\rho^0$ production cross section for real
($Q^2$ = 0) photons.
The results of the fit (solid curves in Fig.~\ref{fig:crossqq2})
are listed in table~\ref{tab:photocross}.

\begin{table} [bt]
\caption{The fitted photoproduction cross sections for $\rho^0$ production
on $^1$H. In the top row the results are shown with the variables
$\sigma_0$ and $\xi^2$ of Eq. (2) as free parameters, the middle row
lists the
results obtained with $\sigma_0$ and $m$ left free, and the bottom
row shows the fitted values of $m$ if a parameterization of the world
data is used to constrain $\sigma_0$ The listed errors include both the
statistical and systematic uncertainties, added in quadrature.
In all cases the $\chi^2$/d.o.f. values of the fits are close to
or less than unity.
\label{tab:photocross}
}
\begin{center}
\begin{tabular}{|c|c|c|}
\hline
Variables & $\langle W \rangle$ = 4.6 GeV & $\langle W \rangle$ = 5.4 GeV \\
\hline
$\sigma_0$ [$\mu$b] & 10.7 $\pm$ 1.8 & 13.1 $\pm$ 1.7 \\
$\xi^2$ & -0.023 $\pm$ 0.049 & -0.13 $\pm$ 0.03 \\
\hline
$\sigma_0$ [$\mu$b] & 11.2 $\pm$ 2.9 & 19.7 $\pm$ 4.9 \\
$m$ & 2.42 $\pm$ 0.18 & 2.83  $\pm$ 0.18 \\
\hline
$\sigma_0$ [$\mu$b] & 11.5 (fixed) & 11.0 (fixed) \\
$m$ & 2.44 $\pm$ 0.05 & 2.46 $\pm$ 0.06 \\
\hline
\end{tabular}
\end{center}
\end{table}

As the ratio $R$ must be positive, the negative fit values of $\xi^2$ indicate
a deficiency of the VMD-model. Hence, a different approach
was also used, i.e. the exponent $m$ of the propagator
in Eq.~(\ref{eq:res-fittingf}) was used as a free parameter, while 
the value of $R(Q^2)$ was fixed using information extracted from 
measurements of the $\rho^0$ decay angular distribution~\cite{belz}:
\begin{equation}
    R(W,Q^2) = c_0(W) \cdot (Q^2 / M_{\rho}^2)^{c_1}.
\label{eq:res-r_q2}
\end{equation}
\noindent
The parameterization of $R$
expressed by Eq.~(\ref{eq:res-r_q2}) has been
obtained from a fit of the world data on $R$, yielding
$c_1$ = 0.61 $\pm$ 0.04, and 
$c_0(W)$ = 0.33 $\pm$ 0.03 (0.48 $\pm$ 0.03) 
for 4 GeV $< W <$ 7 GeV ($W >$ 7 GeV). The
resulting fits (dashed curves in Fig.~\ref{fig:crossqq2}) also
give a good account of the $Q^2$-dependence of the
$\rho^0$ production cross section. 
The fitted parameters are listed in the middle row of
table~\ref{tab:photocross}. The values of $\sigma_0$ are 
somewhat different from the results of the previous method,
but are consistent as the errors have increased considerably.

It is concluded that an increase of the exponent $m$ can avoid
the negative values of $\xi^2$ found in the previous fits. This
result quantitatively confirms similar conclusions obtained by the
E665 collaboration ($m$ = 2.51 $\pm$ 0.07 at $W \approx$ 15
GeV~\cite{e665}).

The fitted values of the extrapolated photoproduction cross section
$\sigma_0$ are in agreement with existing real-photon
data~\cite{desy-rho},
but carry larger error bars particularly if the exponent and
the cross section are treated as free parameters.
Since the VMD cross section scales with $Q^{-4}$ while a 
perturbative QCD description predicts~\cite{Brod} that the longitudinal
part of the cross section scales
with $Q^{-6}$, it is of particular interest to study the
value of the exponent of the propagator in the
transitional domain probed by the present data. By fixing the
value of the photoproduction cross section $\sigma_0$ using
a parameterization of the existing photoproduction
data~\cite{desy-rho},
a one-parameter fit of the $Q^2$-dependence of the cross
section has been carried out, yielding fairly precise values
for the exponent $m$ as can be seen from the bottom row
of table~\ref{tab:photocross}. The fitted value is remarkably
close to 2.5, which is in agreement with results obtained by
E665~\cite{e665} (2.51 $\pm$ 0.07) at an average $W$ value of
17 GeV, but somewhat above the result obtained by 
H1~\cite{H196} (2.24 $\pm$ 0.09) at $W$ = 75 GeV.

%%%%%%%%%%%%%%%%%%%%%%%%%%%%%%%%%%%%%%%%%%%%%%%%%%%%%%%%%%%%%%%%%%%%%%

\begin{figure} [t]
\begin{center}
\includegraphics[width=0.47\textwidth]{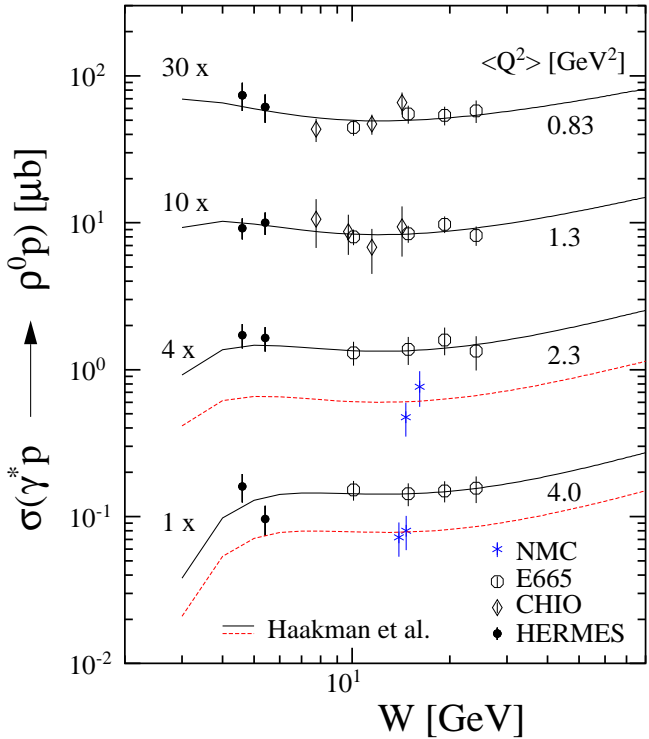}
\caption[]{\label{fig:crossw2}
The virtual-photoproduction cross section for $\rho^0$
production versus $W$ at average $Q^2$ values of 0.83,
1.3, 2.3 and 4.0 GeV$^2$ (from top to bottom).
The coding of the data symbols is the same as in Fig. 2.
The solid (and dashed) lines represent the results of
calculations by L. Haakman {\it et al.}~\cite{ref4}. For the
solid (dashed) lines the HERMES and E665 (NMC) data were used 
to normalize the
curves. Note that both the data and the calculations have
been multiplied by the factors indicated on the left-hand side
of the figure for plotting purposes only.
}
\end{center}
\end{figure}

\subsection{$W$-dependence}

In Fig.~\ref{fig:crossw2} the $\rho^0$ virtual-photoproduction 
cross sections
are shown as a function of $W$ for the four different $Q^2$ bins.
As was argued above, only the data for $W$ $>$ 4 GeV are being
considered. The data are compared to the calculations of
Haakman {\it et al.}~\cite{ref4}, which relate the $W$-dependence of the
vector-meson production cross section to the proton structure
function $F_2^{\mathrm p}(x,Q^2)$. In order to understand how the
$x$-dependence of $F_2^{\mathrm p}(x,Q^2)$ influences the 
$W$-dependence of the $\rho^0$ production cross section, it is
useful to realize that $x$ and $W$ are related by 
$W^2 = Q^2 \frac{1-x}{x} + M^2$, thus showing that the two
variables are essentially inversely proportional at large
values of $Q^2$ and small values of $x$.
The relation between the $\rho^0$ production cross section
and the structure function is apparent in
the expression for the $t$-dependence of the cross
section given in Ref.~\cite{ref4}: 
\begin{equation}
 { \frac{d \sigma}{dt} } = f_{\rho} (Q^2) F_2^{\mathrm p}(x,Q^2)^2 
   \exp[\Lambda(W)t].
\label{eq:haakm}
\end{equation}
In this expression the scaling function $f_{\rho}(Q^2)$ is not
determined by the model, and has been obtained from a fit to the
experimental data displayed in Fig.~\ref{fig:crossw2}.
For $F_2^{\mathrm p}(x,Q^2)$ the parameterization of
Capella {\it et al.}~\cite{Cap} in terms of Pomeron and Reggeon
exchanges has been taken. The slope parameter $\Lambda$ is given by
\begin{equation}
	\Lambda(W) = 2 [ R_{sc}^2 + 2 \alpha_P^{\prime} \ln(W/W_0) ]
\label{eq:haakm2}
\end{equation}
with the radius parameter $R_{sc}^2 = R_{\rm p}^2 + R_{V}^2(Q^2)$ expressing
the combined size of the scattering objects, and $W_0^2$ given
by $M_{\rho^0}^2 + Q^2$. Expressions for $\alpha_P^{\prime}$, the 
slope of the effective Pomeron trajectory, and $R_{V}^2(Q^2)$ are 
given in Ref.~\cite{ref4}.
Thus evaluated, the value of $\Lambda$ turns out to be
consistent with the value of the slope parameter derived
from a fit of the $-t'$ distributions~\cite{ref4}.

In order to compare the $W$-dependence of the model of Ref.~\cite{ref4} to 
that of the data, the calculations have been normalized to the 
HERMES and E665 data, for each $Q^2$ bin separately.
The results are displayed in Fig.~\ref{fig:crossw2} for
$Q^2$ bins centered at 0.83, 1.3, 2.3 and 4.0 GeV$^2$, respectively.
A good description of these data is obtained,
thus confirming the findings of Ref.~\cite{ref4} but now at lower 
values of $W$. It should be noted, however, that the data
collected at $W$ $<$ 4 GeV -- as displayed in Fig.~\ref{fig:crossw1} --
show a steep rise with decreasing $W$ that
is not reproduced by the calculations of Ref.~\cite{ref4}. 
As was mentioned above, this is presumably caused by 
additional reaction processes that are not contained in
the theoretical framework of Ref.~\cite{ref4}. Without
explicit calculations that include these effects, it
cannot be excluded that there is a finite contribution
due to such processes above $W$ = 4 GeV, especially
at the lowest $Q^2$ values.

The NMC data (also shown in Fig.~\ref{fig:crossw2}) can be compared 
to the calculations of Ref.~\cite{ref4} as well. This leads to differently 
normalized curves (dashed lines in Fig.~\ref{fig:crossw2}). If the NMC
normalization is adopted, the curves fall below the E665 and
HERMES data. Within the constraints of the present
model calculation, the E665 normalization is therefore preferred.
However, on the basis of the HERMES data alone no such distinction
can be made.

%%%%%%%%%%%%%%%%%%%%%%%%%%%%%%%%%%%%%%%%%%%%%%%%%%%%%%%%%%%%%%%%%%%%

\subsection{The longitudinal cross section}

The results of recent OFPD calculations,
such as those described in Refs.~\cite{MvdH,MvdH2,Mank1,Mank2},
only concern the longitudinal component $\sigma_L$ of the
$\rho^0$ virtual-photoproduction cross section, because the
factorization theorem~\cite{Collins} applies only to the longitudinal case.
Hence, values of $\sigma_L$ were extracted from the data.

\begin{figure} [t]
\begin{center}
\includegraphics[width=0.47\textwidth]{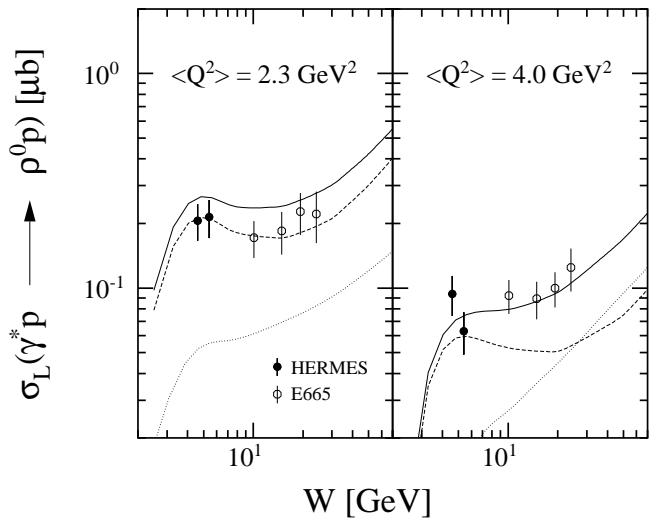}
\caption[]{\label{fig:crossw3}
The longitudinal component of the
virtual-photoproduction cross section for $\rho^0$
production versus $W$ at average $Q^2$ values of
2.3 (left) and 4.0 GeV$^2$ (right).
The solid lines represent the results of
the calculations of Refs.~\cite{MvdH,MvdH2}.
The dashed (dotted) curves represent the quark
(two-gluon) exchange contributions within these
calculations.
}
\end{center}
\end{figure}

The longitudinal cross section $\sigma_L$ is
related to the total $\rho^0$ production cross section 
$\sigma_{\gamma^* p \rightarrow \rho^0 p}$:
\begin{equation}
\sigma_L ={{R} \over {1 + \epsilon R}} \sigma_{\gamma^* p \rightarrow \rho^0 p}.
\label{eq:LT}
\end{equation}
Assuming SCHC the ratio 
$R = \sigma_L / \sigma_T$ can be derived from
the longitudinal fraction $r_{00}^{04}$ of $\rho^0$
mesons, which can be extracted from their polar
decay angular distribution $W(\cos\Theta)$.
In Ref.~\cite{belz} such an analysis is presented,
which has provided us with the parameterization of $R$
shown in Eq.~(\ref{eq:res-r_q2}). This has been used to
evaluate $\sigma_L$ for each of the cross sections listed in
Table~\ref{tab:cross}.
(Note that $\epsilon$ can be determined from the kinematics
of each event.)
The results are listed in Table~\ref{tab:longcross}
and, for $Q^2 >$ 2 GeV$^2$, displayed in Fig.~\ref{fig:crossw3}.
It may be noted that the values of $\sigma_L$ increase
by about 30\% if the quoted parameterization of $R$ for $W$ $>$
7 GeV is used.

The data are compared to the model calculations described in
Ref.~\cite{MvdH}, but with the improvements described in Ref.~\cite{MvdH2}: 
higher twist effects are included
in a phenomenological fashion, the $Q^2$-dependence of the
strong coupling constant $\alpha_s$
was accounted for, and the sea quarks
are included using the MRST98 parton
distributions~\cite{MRST98}. As can be seen from Fig.~17 
of Ref.~\cite{MvdH2} the higher-twist contributions, which were included
through the dependence on the intrinsic transverse 
momentum $k_{\perp}$, reduce the longitudinal cross section
by typically a factor of 5 at $Q^2 \approx$ 2 GeV$^2$. 
As the (phenomenological) higher twist corrections at 
lower $Q^2$ values are even larger,
the comparison between the OFPD calculations of 
Ref.~\cite{MvdH,MvdH2} and the data has not been extended to the 
data below $Q^2 \approx$ 2 GeV$^2$. Moreover, the OFPD
calculations cannot be used for data collected at $W$ $<$ 4 GeV,
as no other reaction channels were included.

\begin{table} [bt]
\caption{Longitudinal cross sections $\sigma_L$ for $\rho^0$ virtual 
photoproduction on $^1$H in $\mu$b. The listed uncertainties include 
both the total
error on the measured $\rho^0$ virtual-photoproduction cross sections
and the error on the parameterization of $R$ for $W$ $<$ 7 GeV, which
was used in the evaluation of Eq.~(\ref{eq:LT}). The average values
of $\epsilon$ for each bin are listed as well.\
\label{tab:longcross}
}
\begin{center}
\begin{tabular}{|c|c|c|c|c|}
\hline
$<Q^2>$ & \multicolumn{2}{c|}{$<W>$ = 4.6 GeV} & 
\multicolumn{2}{c|}{$<W>$ = 5.4 GeV} \\
\hline
[GeV$^2$] & $<\epsilon>$ & $\sigma_L$ [$\mu$b] & $<\epsilon>$ & 
$\sigma_L$ [$\mu$b] \\
\hline
0.83 & 0.87 & 0.77 $\pm$ 0.16 & 0.74 & 0.67 $\pm$ 0.14 \\
1.3  & 0.87 & 0.35 $\pm$ 0.07 & 0.73 & 0.41 $\pm$ 0.08 \\
2.3  & 0.86 & 0.21 $\pm$ 0.04 & 0.71 & 0.21 $\pm$ 0.04 \\
4.0  & 0.83 & 0.09 $\pm$ 0.02 & 0.69 & 0.06 $\pm$ 0.01 \\
\hline
\end{tabular}
\end{center}
\end{table}

The OFPD calculations are compared in Fig.~\ref{fig:crossw3}
to the longitudinal
cross sections measured by E665 \cite{e665} and the
present experiment. The calculated 
contributions due to quark exchange 
(dashed curves) and two-gluon exchange (dotted curves)
are also shown separately.
In the present kinematic domain the quark contributions
dominate, and only at $Q^2$ = 4.0 GeV$^2$
and $W >$ 10 GeV the gluon contribution starts to contribute
significantly. A fairly good agreement between the calculations 
and the data is obtained, given 
the existing theoretical uncertainties related to the size
of the higher-twist contributions, and
the relatively low $Q^2$ values involved.
The calculated rise of the cross section at $W \approx$ 5 GeV, which is 
associated with the contribution due to the exchange of valence quarks,
is not inconsistent with the new HERMES data. 

%%%%%%%%%%%%%%%%%%%%%%%%%%%%%%%%%%%%%%%%%%%%%%%%%%%%%%%%%%%%%%%%%%%%%%%%%%

\section{Summary}\label{sec:sum}
  
In summary, cross sections have been presented for exclusive diffractive
$\rho^0$ virtual photoproduction
in the $W$-domain between 4 and 6 GeV. The
$Q^2$-dependence of the cross section is well described by the
propagator of the Vector Meson Dominance model,
although an increase of the exponent
from its original value of 2 to about 2.5 is needed
when consistency with
existing data for $R = \sigma_L / \sigma_T$ is required. By extrapolating the
$ \gamma^* p \rightarrow \rho^0 p$ cross section to $Q^2$ = 0,
photoproduction cross sections are found which are in fair agreement
with previously published values of about 11 $\mu$b.
The $W$-dependence
of the present and (most) existing data in the $W$-domain between
4 and 25 GeV is well described by a model linking the
$x$-dependence of the proton
structure function $F_2^{\mathrm p}(x,Q^2)$ to the vector meson production
cross section. The longitudinal component of the
cross section has been compared to calculations based on
the Off-Forward Parton Distribution framework. A fairly good agreement
with the data is found.

\begin{acknowledgement}
We gratefully acknowledge the DESY Directorate for its support and
the DESY staff and the staffs of the collaborating institutions.
We particularly appreciate the efforts of the HERA machine group
in providing high beam polarization.
Additional support for this work was provided by
the Deutscher Akademischer Austauschdienst (DAAD) and
INTAS, HCM,  and TMR network contributions from the European Community.
\end{acknowledgement}

\end{document}